\documentclass[twocolumn,showpacs,preprintnumbers,amsmath,amssymb,prb]{revtex4}
\usepackage[dvips]{color,graphics,epsfig,rotating}
\usepackage{graphicx}
\usepackage{dcolumn}
\usepackage{bm}

\begin{document}

\title{Bonding of H in O vacancies of ZnO}

\author{H. Takenaka and D.J. Singh}
\affiliation{Materials Science and Technology Division and
Center for Radiation Detection Materials and Systems, Oak Ridge National
Laboratory, Oak Ridge, Tennessee 37831-6032} 

\date{\today} 

\begin{abstract}
We investigate the bonding of H in O vacancies in ZnO using
density functional calculations. We find that H is anionic
and does not form multicenter bonds with Zn in this compound.
\end{abstract}


\maketitle

ZnO is of importance as an extremely
fast inorganic scintillator material when
doped with Ga or In. It is useful in alpha particle detection, e.g. for
devices such as deuterium-tritium neutron generators used in
radiography.
\cite{lehmann,luckey,batsch,derenzo,neal,bourret}
In this application, H treatment has been shown to improve properties.
ZnO has also attracted much recent attention
motivated by potential applications as an oxide electronic
material,
\cite{hoffman,nomura,suzuki,look-02}
and in optoelectronic and lighting applications.
\cite{nicoll,bagnall,look-99,look,huang}
H has been implicated as playing an important role in the electronic
properties for ZnO for those applications as well.
\cite{vandewalle,cox,hoffmann}
From a fundamental point of view,
the behavior, and especially bonding of H, is of great interest;
H plays an exceptionally important role in chemistry, and
shows unique bonding characteristics. For example, it readily forms
compounds where it behaves as a halogen ion and forms structures similar to
fluorides, such as rutile,
perovskite, rocksalt, etc.,
\cite{bertheville,gingle,yvon}
and at the same time readily occurs
a cation in other chemical
environments. Polar covalent bonds involving H and hydrogen bonds
are central to
much of organic chemistry as well the properties of important
substances such as water.
\cite{pauling-book}
Thus the recent report by Janotti and Van de Walle (JV) that H forms
a new type of strong
multicenter bond in O vacancies in ZnO is of wide ranging interest.
\cite{janotti}

In this paper, we present standard local density approximation (LDA)
calculations of the
electronic properties and structure of H containing O vacancies in ZnO.
We do not find the multicenter covalent bonds claimed by JV,
and instead characterize the behavior of H as quite conventional in that
it occurs as an anion on the anion site in a polar crystalline environment.

Our calculations were done within the standard local density approximation
using the general potential linearized augmented planewave (LAPW) method,
including local orbitals. \cite{singh-book,singh-lo} Specifically,
we constructed a 72 atom 3x3x2 wurtzite supercells of ZnO, with one O
atom removed and replaced by H. The calculations were done using the
bulk lattice parameters of ZnO, but the internal coordinates of all
atoms in the supercell were fully relaxed.
No symmetry was assumed in the relaxations.
The LAPW method is an all electron method that makes no shape approximations
to either the potential or charge density.
It divides space into non-overlapping atom centered spheres and
an interstitial region.
The method then employs accurate basis sets appropriate
for each region. \cite{singh-book}
In the present calculations,
LAPW sphere radii of
2.0 $a_0$, 1.6 $a_0$ and 1.2 $a_0$ were used for Zn, O, and H
respectively, along with a basis set consisting of more than
8500 LAPW functions and local orbitals. Convergence tests were done
with a larger basis set of approximately 12000 functions, but no
significant changes were found. The relaxations were done without
any imposed symmetry, with a 2x2x2 special ${\bf k}$-point zone
sampling. A sampling using only the $\Gamma$ point was found to 
yield slightly different quantitative results, due to the limited size of
our supercell, but would lead to the same conclusions. 
The calculated value of the internal parameter is $u$=0.119, which
agrees almost exactly with the experimental value.
The densities of states used to analyze the electronic
properties were obtained using the linear tetrahedron method based
on eigenvalues and wavefunctions at 36 ${\bf k}$-points in the half
zone ({\bf k} and -{\bf k} are connected by time reversal).

In our relaxed structure for a neutral cell, we find that H occurs
in a slightly asymmetric position, with three Zn neighbors at 2.03 \AA,
and one Zn neighbor (the one along the $c$-axis direction) at 2.17 \AA.
For the singly charged cell, we obtain a very similar result,
specifically three Zn neighbors at 2.02 \AA, and the apical Zn at
2.21 \AA. In the following, we focus on the neutral cell except as
noted.

\begin{figure}[tbp]
\vspace{0.5cm}
\includegraphics[width=3.4in,angle=0]{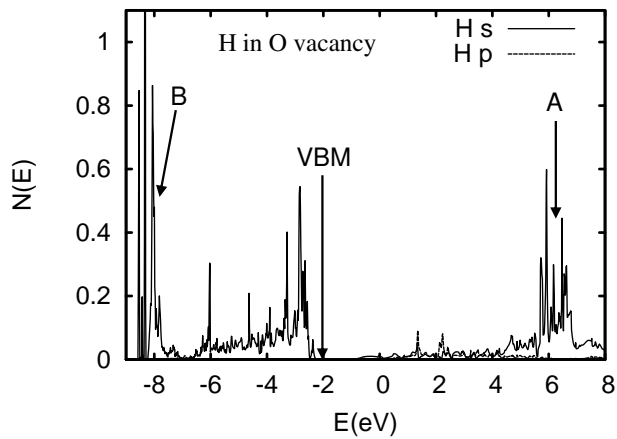}
\caption{Projection of the electronic density of states
onto the $s$ and $p$ components inside the H LAPW sphere, radius
1.2 $a_0$, for
the 72 atom neutral cell.
The two peaks identified by JV as bonding and antibonding
combinations are indicated by ``B" and ``A", respectively. The
Fermi level lies in the conduction bands. The position of the
valence band maximum is denoted by ``VBM" (note that the LDA
strongly underestimates the 3.3 eV band gap of ZnO).
}
\label{dos}
\end{figure}

Fig. \ref{dos} shows the projection of the electronic density
of states onto the H LAPW sphere, of radius 1.2 $a_0$.
The Fermi energy for our neutral cell lies at a position one electron per
cell into the conduction bands, corresponding to the valence difference of
one between O and H.
As may be seen, there are two prominent peaks in the H component of the
density of states, one, denoted 
``B", at $\sim$ -8 eV with respect to the Fermi level
(-6 eV with respect to the valence band maximum), and the other,
denoted ``A", high in
the conduction bands at $\sim$ 6 eV.
JV identified these peaks, ``B"
and ``A", respectively, as the bonding and antibonding combinations of
metal and H orbitals giving rise to the multicenter bond.
In addition, there is significant H $s$ character distributed over
the valence bands, especially near the valence band maximum.
We note that the very large bonding-antibonding splitting of 14 eV
implied by the 
assignment of JV
indicates
extremely strong covalent bonds, which is somewhat surprising
considering the Zn-H distances.
In any case, such a large covalent gap would imply that the bonding
and antibonding states should have mixed character. In other words, the
bonding state should be of roughly half H $s$ character, while
the remaining H 1$s$ character should occur in the unoccupied
antibonding level, so that the occupancy of the H 1$s$ orbital should
be roughly 1 e, and certainly significantly less than 2 e.

\begin{figure}[tbp]
\vspace{0.5cm}
\includegraphics[width=3.4in,angle=0]{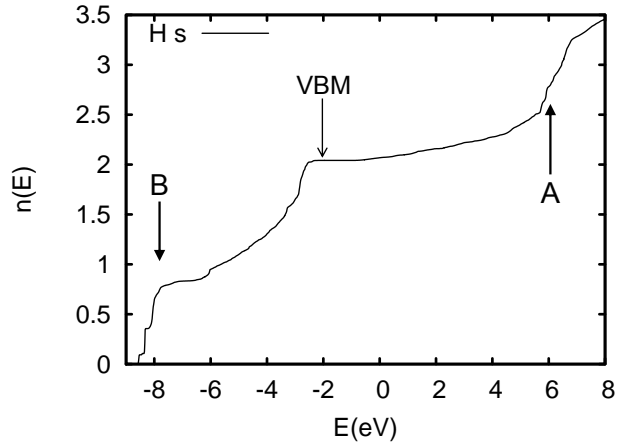}
\caption{Integration of the projection of the H $s$ projected density
of states as in Fig. \ref{dos} normalized according to the
fraction of charge inside a 1.2 $a_0$ sphere for a Watson sphere
stabilized H$^-$ anion (see text).
}
\label{int}
\end{figure}

To analyze the bonding further it is convenient to compare the
charge density with an ionic model, as was done for some alanates.
\cite{aguayo,singh-alanate}
As mentioned, H is known to enter
some solids as an anion, including tetragonal MgH$_2$. \cite{yu}
ZnH$_2$ also exists though it is not as
well characterized. \cite{wiberg}
Furthermore, the simplest
hydride, LiH, is of this ionically
bonded type and includes H$^-$ anions coordinated by six
metal atoms. \cite{kunz,dovesi}
In these hydrogen
anion based materials, the negative H ion is stabilized by the Ewald
field.
In fact, the importance
of the Ewald field is one of the essential differences between
chemistry in solid state and the chemistry of molecules. The
long range Coulomb interaction stabilizes ionic bonding for species
that would generally be largely covalent in small molecules, and
in particular stabilizes anions such as O$^{2-}$
and H$^-$, which are common in solid state chemistry but much less
so in gas phase molecules.
This stabilization by the Ewald field is reflected in the
variability of the effective size of H in
crystal structure data for anionic hydrides. \cite{morris,shannon,pauling}
In view of the common occurrence of H as an anion in many metal hydrides,
it would not
be surprising if H$^{-}$ were stabilized by the
Ewald field of an anion vacancy in a polar
crystal such as ZnO.
Thus we consider an ionic model, based on the charge density of a
H$^-$ ion stabilized by the Ewald field, as simulated by a Watson
sphere, \cite{watson}
as in Ref. \onlinecite{singh-alanate}.
For such a H$^-$ ion, 0.525 e out of 2.0 e, i.e. $\sim$ 26\% of the charge,
is inside a radius of 1.2 $a_0$, so the majority of the charge is outside.
Because of the small sphere radius used for H in our calculations
the amount of charge inside the sphere 
is only weakly dependent on the
Watson sphere radius, which reflects the environment.
For a non-spin polarized neutral H in free space as described in the LDA,
0.378 e (38\%) would be inside a radius of 1.2 $a_0$, showing that there is a 
strong dependence on the charge state, though not precise proportionality.
Fig. \ref{int} shows the integral of the H $s$ character as
a function of energy normalized by the fraction of the H$^-$ charge inside
a 1.2 $a_0$ sphere (0.525/2). Over the valence band region the $p$ contribution
is less than 2\%, and the $d$ contribution is less than 0.2\%.
The conduction bands, which are more Zn $sp$
derived, show a larger proportion of H $p$ character, as may be seen
in Fig. \ref{dos}.
Thus the charge inside the sphere, which comes
from the occupied valence bands, is mainly
due to H $s$ states, and not from orbitals on neighboring atoms.

Using the ionic model for H$^-$, i.e. incorporating the factor of 0.525/2
as the fraction of charge inside the H LAPW sphere, and integrating,
one finds that the peak ``B" contains $\sim$ 0.8 H $s$ electrons. Integrating
over the remaining valence bands brings the H $s$ count to 2.0 electrons,
i.e. what is expected for H$^-$.
This leads to an interpretation of the electronic structure, where the
peak ``B" comes from the H 1$s$ state. This hybridizes
with valence band states, which have mixed Zn $d$ and O $p$ character.
The second peak ``A", 14 eV higher, is then the H $s$ resonance.
This is a very reasonable position for the resonance
of H$^-$. In particular,
the H$^{--}$ resonance of atomic H$^-$ is at $\sim$ 14.5 eV.
\cite{walton,taylor,schulz}
JV emphasized the shape of the charge density associated with the states
in the peak ``B" and argued for bonding based partly on real space images
of this charge. As mentioned,
in our projected density of states we find that this peak
contains 0.8 $s$ electrons (i.e. 40\%
H $s$ character), which would
be consistent with a bonding orbital. However, the hybridization is with
other occupied states, and when the integration is done over all the valence
bands, we find 2 $s$ electrons, consistent with H$^-$. We emphasize that
mixing of occupied states does not contribute to the energy, and that
such hybridizations do not constitute bonds.
Our calculated binding energy relative H$_2$
and a relaxed neutral supercell with an O vacancy is 87 kJ/mol H. \cite{lda-h2}
This may be an overestimate due to LDA errors, \cite{lda}
but in any case is much
smaller than the binding that would be suggested by a 14 eV bonding-antibonding
splitting.

We also calculated the positron wavefunction and lifetimes
for ZnO with an O vacancy and with the H containing O vacancy.
This was done using the LAPW method in the full inverted self-consistent
Coulomb potential plus 
the
correlation and enhancement factors of Boronski and Nieminen
\cite{boronski}
as calculated from the full charge density.
We obtain a bulk positron lifetime for
ZnO of 144 ps, which is
at the lower end of the experimental range.
Reported experimental values are
151 ps (Ref. \onlinecite{bauer}),
170 ps (Ref. \onlinecite{tuomisto}),
141 ps - 155 ps (Ref. \onlinecite{dutta}), and
182 ps (Ref. \onlinecite{chen}).
Significantly, positrons,
which are positively charged, tend to localize in voids and in sites
that are favorable for cations, and localize weakly if at all in anion
sites, due to the unfavorable Coulomb potential. We do not find positron
localization at the O vacancy in our ZnO supercell, indicating that the
O is indeed an anion as expected, nor do we find positron localization or a
significant lifetime increase in the cell with a H containing O vacancy.
We also find no significant change in lifetime for H in an
O vacancy within a charged supercell
with one electron removed.
In contrast, we obtain a bound positron state for Zn vacancies,
both with and without H, reflecting
the fact that Zn is on a cation site.
The calculated lifetime in a supercell with a Zn
vacancy is 212 ps, while with a H filled Zn vacancy we obtain
175 ps (in this case H bonds to a single adjacent O to form a 
hydroxyl like unit with H-O bond length of 1.01 \AA).
\cite{rmt-note}

To summarize, we have performed density
functional calculations for ZnO supercells with both
empty and H filled O vacancies. Based on an analysis of the electronic
structure we do not
find any evidence for hydrogen multicenter bonds, but rather
find that H occurs as H$^-$.

We are grateful for helpful discussions with
L.A. Boatner, J.S. Neal, and L.E. Halliburton.
This work was supported by the Department of Energy,
Office of Nonproliferation Research and Development, NA22.


\begin{references}

\bibitem{lehmann}
W. Lehmann, Solid State Electronics {\bf 9}, 1107 (1966).

\bibitem{luckey}
D. Luckey, Nucl. Instr. and Meth. {\bf 62}, 119 (1968).

\bibitem{batsch}
T. Batsch, B. Bengtson, and M. Moszynski,
Nucl. Instr. and Meth. {\bf 125}, 443 (1975).

\bibitem{derenzo}
S.E. Derenzo, E. Bourret-Courchesne, M.J. Weber, and M.K. Klintenberg,
Nucl. Instr. Meth. Phys. Res. A {\bf 537}, 261 (2005).

\bibitem{neal}
J.S. Neal, L.A. Boatner, N.C. Giles, L.E. Halliburton, S.E. Derenzo,
and E.D. Bourret-Courchesne,
Nucl. Inst. Meth. Phys. Res. A {\bf 568}, 803 (2006).

\bibitem{bourret}
E.D. Bourret-Courchesne, and S.E. Derenzo,
2006 IEEE Nuclear Science Symposium Conference Record N40-5, 1541
(2006).

\bibitem{nomura}
K. Nomura, H. Ohta, K. Ueda, T. Kamiya, M. Hirano, and H. Hosono,
Science {\bf 300}, 5623 (2003).

\bibitem{hoffman}
R.L. Hoffman, B.J. Norris, and J.F. Wager,
Appl. Phys. Lett. {\bf 82}, 733 (2003).

\bibitem{suzuki}
A. Suzuki, T. Matsushita, T. Aoki, Y. Yoneyama, and M. Okuda,
Jpn. J. Appl. Phys., Part 2 {\bf 38}, L71 (1999).

\bibitem{look-02}
D.C. Look, D.C. Reynolds, C.W. Litton, R.L. Jones, D.B. Eason, and G. Cantwell,
Appl. Phys. Lett. {\bf 81}, 1830 (2002).

\bibitem{nicoll}
F.H. Nicoll, Appl. Phys. Lett. {\bf 9}, 13 (1966).

\bibitem{bagnall}
D.M. Bagnall, Y.F. Chen, Z. Zhu, T. Yao, S. Koyama, M.Y. Shen, and T. Goto,
Appl. Phys. Lett. {\bf 70}, 2230 (1997).

\bibitem{look-99}
D.C. Look, J.W. Hemsky, and J.R. Sizelove,
Phys. Rev. Lett. {\bf 82}, 2552 (1999).

\bibitem{look}
D.C. Look, Mater. Sci. Eng. B {\bf 80}, 383 (2001).

\bibitem{huang}
M.H. Huang, S. Mao, H. Feick, H.Q. Yan, Y.Y. Wu, H. Kind, E. Weber,
R. Russo, and P. Yang, Science {\bf 292}, 1897 (2001).

\bibitem{vandewalle}
C.G. Van de Walle, Phys. Rev. Lett. {\bf 85}, 1012 (2000).

\bibitem{cox}
S.F.J. Cox, E.A. Davis, S.P. Cottrell, P.J.C. King, J.S. Lord,
J.M. Gil, H.V. Alberto, R.C. Vilao, J. Pironto Duarte,
N. Ayres de Campos, A. Weidinger, R.L. Lichti, and S.J.C. Irvine,
Phys. Rev. Lett. {\bf 86}, 2601 (2001).

\bibitem{hoffmann}
D.M. Hoffmann, A. Hofstaetter, F. Leiter, H. Zhou, F. Henecker,
B.K. Meyer, S.B. Orlinskii, J. Schmidt, and P.G. Baranov,
Phys. Rev. Lett. {\bf 88}, 045504 (2002).

\bibitem{bertheville}
B. Bertheville, T. Herrmannsdorfer, and K. Yvon,
J. Alloys Compd. {\bf 325}, L13 (2001).

\bibitem{gingle}
F. Gingle, T. Vogt, E. Akiba, and K. Yvon,
J. Alloys Compd. {\bf 282}, 125 (1999).

\bibitem{yvon}
K. Yvon, Chimia {\bf 52}, 613 (1998).

\bibitem{pauling-book}
L. Pauling, {\em Nature of the Chemical Bond} (Cornell University Press,
Ithaca, 1960).

\bibitem{janotti}
A. Janotti, and C.G. Van de Walle, Nature Materials {\bf 6}, 44 (2007).

\bibitem{singh-book}
D.J. Singh and L. Nordstrom, {\em Planewaves, Pseudopotentials and the
LAPW Method, 2nd. Ed.} (Springer, Berlin, 2006).

\bibitem{singh-lo}
D. Singh, Phys. Rev. B {\bf 43}, 6388 (1991).

\bibitem{yu}
R. Yu and P.K. Lam, Phys. Rev. B {\bf 15}, 8730 (1988).

\bibitem{morris}
D.F.C. Morris and G.L. Reed, 
J. Inorg. Nucl. Chem. {\bf 27}, 1715 (1965).

\bibitem{shannon}
R.D. Shannon, Acta Cryst.{\bf A32}, 751 (1976).

\bibitem{pauling}
L. Pauling, Acta Cryst., Sect. B {\bf 34}, 746 (1978).

\bibitem{aguayo}
A. Aguayo and D.J. Singh, Phys. Rev. B {\bf 69}, 155103 (2004).

\bibitem{singh-alanate}
D.J. Singh, Phys. Rev. B {\bf 71}, 216101 (2005).

\bibitem{wiberg}
E. Wiberg, W. Henle, and R. Bauer,
Z. Naturforsch. B {\bf 6}, 393 (1951).

\bibitem{kunz}
A.B. Kunz and D.J. Mickish, Phys. Rev. B {\bf 11}, 1700 (1975).

\bibitem{dovesi}
R. Dovesi, C. Ermond, E. Ferrero, C. Pisani, and C. Roetti,
Phys. Rev. B {\bf 29}, 3591 (1984).

\bibitem{watson}
R.E. Watson, Phys. Rev. {\bf 111}, 1108 (1958); we used a H anion stabilized
by a sphere of radius 1.62 \AA.

\bibitem{walton}
D.S. Walton, B. Peart, and K. Dolder, J. Phys. B {\bf 3}, L148 (1970).

\bibitem{taylor}
H.S. Taylor and L.D. Thomas, Phys. Rev. Lett. {\bf 28}, 1091 (1972).

\bibitem{schulz}
G.J. Schulz, Rev. Mod. Phys. {\bf 45}, 378 (1973).

\bibitem{lda-h2}
The value used for the energy of H$_2$ is -2.294 Ry; 
D.J. Singh, M. Gupta, and R. Gupta, Phys. Rev. B {\bf 75}, 035103 (2007).

\bibitem{lda}
The LDA generally overbinds solids, and this leads to overestimates
of the binding of H in solids, typically in the range of 0 to 20 kJ/mol H;
H. Smithson, C.A. Marianetti, D. Morgan, A. Van der Ven, A. Predith,
and G. Ceder, Phys. Rev. B {\bf 66}, 144107 (2002);
S.V. Halilov, D.J. Singh, M. Gupta, and R. Gupta, Phys. Rev. B {\bf 70}, 
195117 (2004);
K. Miwa and A. Fukumoto, Phys. Rev. B {\bf 65}, 155114 (2002).

\bibitem{boronski}
E. Boronski and R.M. Nieminen, Phys. Rev. B {\bf 34}, 3820 (1986);
see also M.J. Puska and R.M. Nieminen, Rev. Mod. Phys. {\bf 66}, 841 (1994);
P. Schultz and K.G. Lynn, Rev. Mod. Phys. {\bf 60}, 701 (1988).

\bibitem{bauer}
G. Bauer, W. Anwand, W. Skorupa, J. Kuriplach, O. Melikhova, C. Moisson,
W. von Wenckstern, H. Schmidt, M. Lorenz, and M. Grundmann,
Phys. Rev. B {\bf 74}, 045208 (2006).

\bibitem{tuomisto}
F. Tuomisto, V. Ranki, K. Saarinen and D.C. Look,
Phys. Rev. Lett. {\bf 91}, 205502 (2003).

\bibitem{dutta}
S. Dutta, M. Chakrabarti, S. Chattopadhyay, D. Jana, D. Sanyal, and A. Sarkar,
J. Appl. Phys. {\bf 98}, 053513 (2005).

\bibitem{chen}
Z.Q. Chen, S. Yamamoto, M. Maekawa, A. Kawasuso, X.L. Yuan, and
T. Sekiguchi, J. Appl. Phys. {\bf 94}, 4807 (2003).

\bibitem{rmt-note}
The calculations for H in a Zn vacancy were done using a different
set of LAPW sphere radii, as was necessary due to the short H-O
bond length.

\end{references}
\end{document}